# Cheating is evolutionarily assimilated with cooperation in the continuous snowdrift game

Tatsuya Sasaki[1,2,*] and Isamu Okada[3,4]

April 6, 2015

[1]Faculty of Mathematics, University of Vienna, 1090 Vienna, Austria

[2]Evolution and Ecology Program, International Institute for Applied Systems Analysis (IIASA), 2361 Laxenburg, Austria

[3]Department of Business Administration, Soka University, 192-8577 Tokyo, Japan

[4]Department of Information Systems and Operations, Vienna University of Economics and Business, 1020 Vienna, Austria

*Author for correspondence:

Faculty of Mathematics, University of Vienna, Oskar-Morgenstern-Platz 1, 1090 Vienna, Austria

+43 1 4277 50774

tatsuya.sasaki@univie.ac.at

**Highlights**

- We fully analyze continuous snowdrift games with quadratic payoff functions in diversified populations
- It is well known that classical snowdrift games maintain the coexistence of cooperators and cheaters
- We clarify that the continuous snowdrift games often lead to assimilation of cooperators and cheaters
- Allowing the gradual evolution of cooperative behavior can facilitate social inequity aversion in joint ventures

**Abstract.** It is well known that in contrast to the Prisoner's Dilemma, the snowdrift game can lead to a stable coexistence of cooperators and cheaters. Recent theoretical evidence on the snowdrift game suggests that gradual evolution for individuals choosing to contribute in continuous degrees can result in the social diversification to a 100% contribution and 0% contribution through so-called evolutionary branching. Until now, however, game-theoretical studies have shed little light on the evolutionary dynamics and consequences of the loss of diversity in strategy. Here we analyze continuous snowdrift games with quadratic payoff functions in dimorphic populations. Subsequently, conditions are clarified under which gradual evolution can lead a population consisting of those with 100% contribution and those with 0% contribution to merge into one species with an intermediate contribution level. The key finding is that the continuous snowdrift game is more likely to lead to assimilation of different cooperation levels rather than maintenance of diversity. Importantly, this implies that allowing the gradual evolution of cooperative behavior can facilitate social inequity aversion in joint ventures that otherwise could cause conflicts that are based on commonly accepted notions of fairness.

**Keywords:** evolution of cooperation; snowdrift game; replicator dynamics; adaptive dynamics; evolutionary branching; speciation in reverse

## 1. Introduction

In daily life, cooperative behavior in joint ventures is a fundamental index that represents the real state of human sociality and is often a matter of degree that can continuously vary and diverge within a wide range. In general, understanding the origin and dynamics of diversity or heterogeneity has been one of the most challenging hot topics in biology and the social sciences (Axelrod, 1997; MaCann, 2000; Valori et al., 2012). However, most traditional game-theoretical studies on cooperation have described the degree of cooperation in terms of discrete strategies, such as cooperators who contribute all and cheaters who do nothing. Compared with matrix games for finite discrete strategies, games for infinite continuous strategies have been relatively little studied (Brännström et al., 2011; Cressman et al., 2012; Le Galliard et al., 2005; Hilbe et al., 2013; Killingback and Doebeli, 2002; Killingback et al., 1999; McNamara et al., 2008; Nakamura and Dieckmann, 2009; Roberts and Sherratt, 1998; Wahl and Nowak, 1999a, 1999b). We should note that a common motivation among previous game-theoretical models with continuous strategies was to resolve the fundamental question, "How altruistic should one be?" (Roberts and Sherratt, 1998).

Crucially, in the last decade it has been clarified that even without specific assortment, very small, occasional mutations on the degree of cooperation can lead subpopulations of the cooperators and cheaters to gradually dissimilate each other out of a uniform population ("evolutionary branching") (Brännström and Dieckmann, 2005; Brown and Vincent, 2014; Doebeli et al., 2004; Parvinen, 2010). This divergence scenario for the cooperation level has been termed the "tragedy of the commune" (Doebeli et al., 2004). Gradual evolution can favor such a state in which a sense of fairness may be minimized, rather than a state in which all adopt the same cooperation level. To date, theoretical and numerical investigations have shown the conditions under which evolutionary branching occurs at the cooperation level, and by also considering ecological dynamics, where even extinction at the population level can follow (Parvinen, 2010, 2011).

Importantly, previous studies implicitly indicated that a heterogeneous population of cooperators and cheaters may be unstable when considering a small mutation (Brown and Vincent, 2014; Doebeli et al., 2004). To the best of our knowledge, this issue has never been seriously tackled, despite the fact that the coexistence of cooperators and cheaters is one of most elementary equilibria in classical $2\times 2$ matrix games as described in Eq. (1) and is also common in nature and human societies. Indeed, little is known about how continuous investment in joint ventures affects what the traditional framework of a two-person symmetric game with two strategies has so far predicted (Doebeli et al., 2013; Shutters, 2013; Tanimoto, 2007; Zhong et al., 2012).

To address this issue, we consider the snowdrift game (Chen and Wang, 2010; Doebeli and Hauert, 2005; Gore et al., 2009; Hauert and Doebeli, 2004; Kun et al., 2006; Maynard Smith, 1982; Sugden, 1986), which has traditionally been a mathematical metaphor to understand the evolution of cooperation, and in particular, it can result in the coexistence of cooperation and cheating or inter-species mutualism (Fujita et al., 2014; Gore et al., 2009; Kun et al., 2006). (The snowdrift game is also well recognized as the chicken or hawk-dove game (Maynard Smith, 1982)). The classical snowdrift game for cooperators and cheaters has been featured by the rank ordering of the four payoff values: $T > R > S > P$ (Doebeli and Hauert, 2005; Sugden, 1986), which are given in the $2\times 2$ payoff matrix for cooperation (C) and cheating (or defection) (D),

$$\begin{array}{cc} & \begin{array}{cc} \mathrm{C} & \mathrm{D} \end{array} \\ \begin{array}{c} \mathrm{C} \\ \mathrm{D} \end{array} & \begin{pmatrix} R & S \\ T & P \end{pmatrix} \end{array} . \tag{1}$$

We note that if $P$ and $S$ have the other order: $P > S$, then this matrix represents the well-known Prisoner's Dilemma, leading to mutual cheating (D-D) (Axelrod and Hamilton, 1981). The rank ordering for the snowdrift game indicates that when starting with the D-D state where both cheat, for one cheater to switch to cooperation is beneficial to both, yet not so is then for the other to switch to cooperation. The following situation may be useful as an example: the front porch of an apartment has been covered by a snowdrift, such that getting out requires someone to shovel the snowdrift. The situation becomes a sort of snowdrift game if a resident is willing to shovel snow and how much snow (C), and a best response for the other resident(s) is to shovel less (or nothing) (D). Considering that shoveling time and effort can vary continuously, this would naturally evoke a question of "How much would high- and low-contributors differ from each other in the snowdrift game?"

In Sect 2, we extend the discrete snowdrift game to continuous cooperation. Figure 1 presents an overview encompassing evolutionary scenarios in the classical and continuous snowdrift games. In Sect 3, we then investigate the gradual evolution of cooperation with small mutations. In the continuous extension we consider quadratic payoff functions for interpolating these four payoff values in Eq. (1). It is known that the continuous model with quadratic payoff functions is at minimum, required for full coverage of basic adaptive dynamics for a population monomorphic with the same level of cooperation (Brown and Vincent, 2008; Doebeli et al., 2004) (see also (Boza and Számadó, 2010; Chen et al., 2012; Zhang et al., 2013) for effects of more generalized payoff functions). We show that adaptive dynamics in the snowdrift game analytically provides a solution whether a population is monomorphic or dimorphic. Finally, in Sect. 4 we provide a summary and discuss the model, results, and future work.

## 2. Models and methods

**(a) Replicator dynamics for cooperators and cheaters**

We consider the $2\times 2$ matrix game in Eq. (1) in infinitely large populations without any assortment. We denote by $P_{\mathrm{C}}(n)$ and $P_{\mathrm{D}}(n)$ the expected payoffs for a cooperator (C) and cheater (D), respectively, in the population with the frequency of cooperators $n$. Clearly,

$$\begin{aligned} P_{\mathrm{C}}(n) &= nR+(1-n)S, \\ P_{\mathrm{D}}(n) &= nT+(1-n)P. \end{aligned} \tag{2}$$

We analyze the replicator equation for the frequency of cooperators $n$ (Cressman and Tao, 2014; Hofbauer and Sigmund, 1998),

$$\frac{dn}{dt} = n(P_{\mathrm{C}}(n)-\overline{P}(n)), \tag{3}$$

where $\overline{P}(n) = nP_{\mathrm{C}}(n)+(1-n)P_{\mathrm{D}}(n)$ denotes the average payoff over the population. Equation (3) can be rewritten as

$$\begin{aligned} \frac{dn}{dt} &= n(1-n)(P_{\mathrm{C}}(n)-P_{\mathrm{D}}(n)) \\ &= n(1-n)[n(R-T)+(1-n)(S-P)]. \end{aligned} \tag{4}$$

Therefore, the replicator dynamics in the $2\times 2$ matrix game in Eq. (1) are classified into four types by the sign combination of $S-P$ and $R-T$ (Table 1 and Fig. 2(x)) (Lambert et al., 2014; Santos et al., 2012; Shutters, 2013). In particular, if and only if $S-P>0$ and $R-T<0$ hold, the dynamics have a stable interior equilibrium with

$$n = \frac{S-P}{(S-P)-(R-T)} =: \hat{n}\,. \tag{5}$$

**(b) Continuous snowdrift game with quadratic payoff functions**

We then turn to the continuous snowdrift game (Brown and Vincent, 2008; Doebeli et al., 2004; McNamara et al., 2008; Zhong et al., 2008; Zhong et al., 2012). Each player in a random-matching pair of players has an option to make some investment, which can continuously vary between 0 and $x_{\max}$ with $x_{\max} > 0$, to a joint venture. Provided that the focal player invests $x$ and its opponent, $y$, each will receive the benefit $B(x+y)$ with respect to the accumulated investment $x+y$. The benefit is subtracted by the cost $C(x)$ which depends only on the focal player's investment $x$. Thus, the individual net payoff from the one-shot pairwise interaction is $B(x+y)-C(x)$.

We extend the $2\times 2$ matrix game so that the four components of the matrix are described, respectively, by the values of $B(x+y)-C(x)$ with the extreme levels of investment. We assume that a cheater (D) invests $x=0$ and a cooperator (C) invests $x=x_{\max}$, with $x_{\max}=1$ for simplicity. It is straightforward that the traditional payoff matrix is described as

$R=B(2)-C(1)$, $T=B(1)-C(0)$, $S=B(1)-C(1)$, and $P=B(0)-C(0)$. (6)

In the following we assume that the payoff function is quadratic as $B(x+y)=b_2(x+y)^2+b_1(x+y)$ and $C(x)=c_2x^2+c_1x$. Thus, $B(0)=0$ and $C(0)=0$. This reflects a plausible situation in which no contribution results in no benefit and no cost. Using Eqs. (2) and (6),

$R=4b_2+2b_1-c_2-c_1$, $T=b_2+b_1$, $S=b_2+b_1-c_2-c_1$, and $P=0$. (7)

To fully adhere to the order of $T>R>S>P$, in addition to both inequalities: for $T>R$,

$$R-T=3b_2+b_1-c_2-c_1<0, \tag{8}$$

and, for $S>P$,

$$S-P=b_2+b_1-c_2-c_1>0, \tag{9}$$

it is required that $R>S$, namely,

$$R-S=3b_2+b_1>0. \tag{10}$$

Eqs. (8) and (9) yield that $b_2 < 0$: the quadratic benefit function for the snowdrift game should be concave.

**(c) Monomorphic adaptive dynamics and evolutionary branching**

We are interested in understanding how the strategy distribution over the population changes through imitation of the successful strategies of others (namely, social learning) with small mutations in the continuous snowdrift game. We thus investigate this by means of adaptive dynamics (Deng and Chu, 2011; Geritz et al., 1997; Geritz et al., 1998). The expected payoff for a rare mutant with investment level $y$ among the residents with an investment level $x$ is $P(x,y) = B(x+y) - C(y)$. In the case $x = y$, $P(x,x) = B(2x) - C(x) =: \overline{P}(x)$, represents the average payoff over the monomorphic population with $x$. The growth rate of the rare mutant is the so-called invasion fitness, given by $S(x,y) = P(x,y) - \overline{P}(x)$ in the resident monomorphic population with $x$. We consider $D(x) = \partial_y S(x,y)\big|_{y=x}$ which expresses the selection gradient of the mutant-fitness landscape at $x$. Let $\mu(x)$, $\sigma^2(x)$, and $\hat{m}(x)$ denote the mutation probability, mutation variance, and equilibrium-population size at $x$, respectively. Adaptive dynamics for a monomorphic population with $x$ is governed by the canonical equation $dx/dt = (1/2)\mu(x)\sigma^2(x)\hat{m}(x)D(x)$, except around a singular strategy, $x = \hat{x}$, at which the selection gradient $D(x)$ vanishes. One can set $(1/2)\mu(x)\sigma^2(x)\hat{m}(x)$ to 1 without loss of generality (Meszéna et al., 2001).

In the continuous snowdrift game with quadratic cost and benefit functions, we can use known results (Brännström et al., 2011; Brown and Vincent, 2008; Doebeli et al., 2004). The invasion fitness in the model can be rewritten as

$$S(x,y) = (y-x)[b_2(3x+y) + b_1 - c_2(x+y) - c_1]. \quad (11)$$

Then,

$$D(x) = (4b_2 - 2c_2)x + b_1 - c_1 \quad (12)$$

Thus, there exists at most one singular strategy, $\hat{x}$, given by

$$D(\hat{x}) = 0 \Leftrightarrow \hat{x} = -\frac{b_1 - c_1}{4b_2 - 2c_2}. \quad (13)$$

From $D'(x) = 2(2b_2 - c_2)$, we know that, in a case that $0 < \hat{x} < 1$, it is (convergence) stable, if $2b_2 - c_2 < 0$; otherwise, it is unstable. Moreover, according to the theory of adaptive dynamics, the curvature of invasion fitness at a singular strategy lets us know whether the evolution of populations will end at the singular strategy. In the model, the curvature is given by $\partial_y^2 S(x, y)\big|_{y=x=\hat{x}} = 2(b_2 - c_2)$. Indeed, the singular strategy $\hat{x}$ is evolutionarily stable so that the population at $\hat{x}$ cannot be invaded by any rare mutant neighbors, if invasion fitness takes a maximum at $\hat{x}$, $b_2 - c_2 < 0$; otherwise (if it takes a minimum at $\hat{x}$), it is evolutionarily unstable so that the population at $\hat{x}$ will undergo disruptive selection to a couple of diverging subpopulations.

Therefore, a necessary condition for the interior singular strategy $\hat{x}$ to be convergence stable and evolutionarily unstable (namely, an evolutionary-branching point) is that $2b_2 < c_2 < b_2$. Considering $b_2 < 0$, this yields that evolutionary branching also requires a concave (decelerating) cost function with $c_2 < 0$. Therefore, a convex (accelerating) cost function with $c_2 > 0$ will never result in evolutionary branching (Fig. 2(a)).

**(d) Individual-based models**

For the sake of comparison of results in large, but finite populations, we also considered an existing individual-based model for the continuous snowdrift game (Doebeli et al., 2004). In the model, we iteratively apply the replicator dynamics to finite populations as follows: first, a focal individual $i$ and another individual $j$ are selected at random. Their respective payoffs, $P(i)$ and $P(j)$, are determined independently after giving each of the two individuals a single offer to participate in a public good game. If the focal individual has the lower payoff of the two, i.e., $P(j) > P(i)$, it imitates individual $j$ 's strategy with a probability proportional to the payoff difference $P(j) - P(i)$. Second, independent mutations occur in the focal individual's cooperative investment $x$, each with a small probability $\mu$. If a mutation occurs, the focal individual's new value of cooperative investment is drawn from a normal distribution with standard deviation $\sigma$, centered at its pre-mutational trait value.

## 3. Results

### (a) Coordinate evolutionary outcomes of discrete and continuous snowdrift games

For parameterization, subsequently, we represent the coordinate system with $d_1 = S - P$ and $d_2 = R - T$. Considering Eqs. (8) and (9), hence, $b_1 = (3d_1 - d_2)/2 + c_2 + c_1$ and $b_2 = (d_2 - d_1)/2$. Using the parameter space $(d_1, d_2, c_1, c_2)$, we can overlay classification diagrams of evolutionary scenarios for discrete and continuous strategies (Fig. 2). We note that the coordinate system $(d_1, d_2)$ is equivalent with $(D_g = T - R, D_r = P - S)$ which was originally reported by Tanimoto and Sagara (2007) and has been commonly shared by following application (e.g., Tanimoto, 2007; Zhong et al., 2012; Tanimoto, 2014). For simplicity, in what follows we assume that the values of $c_1$ and $c_2$ are fixed.

We then turn to adaptive dynamics in the continuous snowdrift game. It follows from Eq. (12) that $D(0) = (3d_1 - d_2)/2 + c_2$ and $D(1) = (3d_2 - d_1)/2 - c_2$. The selection gradient $D(x)$ is linear. One can thus describe the full classification of the monomorphic adaptive dynamics in the continuous snowdrift game (Doebeli et al., 2004) in terms of the signs of $D(0)$ and $D(1)$ (Table 2, see Sasaki et al. (Unpublished results) for continuous public good games). In a case where $D(0)$ and $D(1)$ have the same sign ((i) positive or (iii) negative), there is no point at which $D(x)$ vanishes, and therefore, the population unilaterally evolves to (i) $x = 1$ or (iii) $x = 0$, respectively. For case (ii), $D(0) < 0$ and $D(1) > 0$, there is exactly one singular strategy, which is evolutionarily repelling (not convergence stable) and which divides the strategy space into two basins of attraction for maximal investment $x = 1$ and no investment $x = 0$. For case (iv), $D(0) > 0$ and $D(1) < 0$, there is again exactly one singular strategy, which is evolutionarily attracting (convergence stable).

Given a fixed $c_1$ and $c_2$, the intersection of lines $D(0) = 0$ and $D(1) = 0$ in the $(d_1, d_2)$-space is $\mathrm{P} = (-c_2/2, c_2/2)$. In the case, $c_2 < 0$, it follows that point P is located in the fourth quadrant, around which all adaptive scenarios in Table 2 are possible. In contrast to this, having an accelerating cost with $c_2 > 0$ leads to convergence and an evolutionarily stable singular point for all points in the fourth quadrant (the region (iv-A) in Fig. 2).

For cases (ii) or (iv), depending on the curvature of $D(x)$, the population state with the singular strategy can either be (A) evolutionarily stable or (B) unstable. For instance, the combination of (iv) and (B) means that monomorphic populations lead to evolutionary diversification into a mixture of full and no cooperation, entitled (iv-B) evolutionary branching.

Finally, with adhering situations under the social dilemma (Dawes, 1980), the corresponding natural conditions are that $C(x)$ and $B(x)$ are strictly increasing within these domains $[0,1]$ and $[0,2]$, respectively. This requires that $C'(0)=c_2>0$ and $C'(1)=2c_2+c_1>0$, and

$$B'(0)=\frac{3d_1-d_2}{2}+c_2+c_1>0, \tag{14}$$

$$B'(2)=\frac{3d_2-d_1}{2}+c_2+c_1>0. \tag{15}$$

Considering $c_2+c_1>0$, it follows that in the quadrant for snowdrift games only, Eq. (15) matters (and Eq. (14) holds for all $(d_1,d_2)$ in the forth quadrant). We note that $B'(2)>0$ leads to $R-S>0$.

**(b) Classify replicator dynamics for intermediate strategies**

We exclusively analyzed the replicator dynamics for two strategies generally given by $0\le x_2<x_1\le 1$ in the continuous snowdrift game. We denoted a dimorphic population with these strategies as $X=\{(x_1,n_1),(x_2,n_2)\}$, where $n_i$ represents the frequency of $x_i$ for $i=1,2$ (thus, $n_2=1-n_1$). The expected payoff for strategy $x_i$ for $i=1,2$, then was defined by $P(X,x_i)=n_1B(x_1+x_i)+(1-n_1)B(x_2+x_i)-C(x_i)$. We also denoted by $\bar{P}(X):=n_1P(X,x_1)+(1-n_1)B(X,x_2)$ the average payoff over the dimorphic population.

The replicator dynamics for $x_1$'s frequency $n_1$ is

$$\begin{aligned}\frac{dn_1}{dt}&=n_1(P(X,x_1)-\bar{P}(X))\\&=n_1(1-n_1)(P(X,x_1)-P(X,x_2)),\end{aligned} \tag{16}$$

where

$$\begin{aligned} &P(X,x_1)-P(X,x_2) \\ &=(x_1-x_2)[2b_2(x_1-x_2)n_1+b_2(3x_2+x_1)+b_1-c_2(x_2+x_1)-c_1] \\ &=: F_{12}(n_1). \end{aligned} \tag{17}$$

From its linearity, the evolution of $n_1$ is determined by the signs of $F_{12}(0)$ and $F_{12}(1)$. Considering Eq. (11) yields that $F_{12}(0)=S(x_2,x_1)$ and $F_{12}(1)=-S(x_1,x_2)$. That is, the sign pair of $(S(x_2,x_1),S(x_1,x_2))$ (Table 3) can indicate the evolutionary outcome from the replicator dynamics. Therefore, the four criteria (I)-(IV) for classifying the replicator dynamics for D ($x=0$) and C ($x=1$) (Table 1) can similarly be applied to any pair of $x_1$ and $x_2$ on the strategy space $[0,1]$ (Table 3). In particular, for the cases of (II) and (IV), solving $P(X,x_1)-P(X,x_2)=0$ with respect to $n_1$ leads to a non-trivial equilibrium, in which two strategies coexist. The equilibrium frequency is uniquely given by

$$\hat{n}_1(x_1,x_2)=\frac{-b_2(x_1+3x_2)-b_1+c_2(x_1+x_2)+c_1}{2b_2(x_1-x_2)}, \tag{18}$$

as in the supplement of (Doebeli et al., 2004).

We note that in the model, invasion fitness has already been resolved into two linear components: one variable as $b_2(3x+y)+b_1-c_2(x+y)-c_1$ and a fixed diagonal as $y-x$. This leads to the so-called pairwise invasibility plot (PIP) (Geritz et al., 1997; Geritz et al., 1998), a sign plot of invasion fitness $S(x,y)$ on $(x,y)$-space, which can be separated by lines (Fig. 3). The PIP diagram can provide a useful overview to determine the sign pair for any $(S(x_2,x_1),S(x_1,x_2))$ and thus the replicator dynamics in any dimorphic population. The adaptive dynamics of the population, once degenerated to monomorphism, can then be predicted by the four adaptive dynamics criteria in Table 2. In certain cases its dimorphism is protected, otherwise, we shall consider adaptive dynamics in dimorphic populations.

**(c) Dimorphic adaptive dynamics and evolutionary merging**

Previous studies have calculated adaptive dynamics for dimorphic populations when $2b_2 < c_2 < b_2$: the singular strategy is evolutionary-branching. We shall show that in the case of $b_2 - c_2 < 0$, the dimorphism is destabilized and a reverse process of adaptive diversification can occur: the extreme strategies, $x = 1$ and $x = 0$, can evolve towards the interior singular strategy $x = \hat{x}$ with $0 < \hat{x} < 1$. (See Fig. 4 for individual-based simulations.)

We consider adaptive dynamics for dimorphic populations with distribution $X = \{(x_1, n_1), (x_2, n_2)\}$. The expected payoff for a rare mutant with $y$ is then defined by

$$P(X, y) = n_1 B(x_1 + y) + (1 - n_1) B(x_2 + y) - C(y). \quad (19)$$

The invasion fitness for the mutant is given by $S(X, y) = P(X, y) - \overline{P}(X)$ (Geritz et al., 1997). For the quadratic cost and benefit functions, the adaptive dynamics for the dimorphic population $X$ with $0 \le x_2 < \hat{x} < x_1 \le 1$, are given by

$$\begin{aligned} \dot{x}_1 &= m_1(x_1, x_2)\partial_y S(X, y)\Big|_{y=x_1} = m_1(x_1, x_2)(b_2 - c_2)(x_1 - x_2), \\ \dot{x}_2 &= m_2(x_1, x_2)\partial_y S(X, y)\Big|_{y=x_2} = -m_2(x_1, x_2)(b_2 - c_2)(x_1 - x_2), \end{aligned} \quad (20)$$

where $m_1$ and $m_2$ are positive quantities that describe the mutation process in the two branches with $x_1$ and $x_2$; and, $m_1$ and $m_2$ are proportional to $n_1$ and $1 - n_1$, respectively (Doebeli et al., 2004; Meszéna et al., 2001).

In a case where $b_2 - c_2 = (d_2 - d_1 - 2c_2)/2 > 0$, as shown in Doebeli et al. (2004), it follows that $\dot{x}_1 > 0$ and $\dot{x}_2 < 0$, and thus, the two branches are repelling each other. We note that the PIP in Fig. 3(d) indicates that for all of two strategies with $0 \le x_2 < \hat{x} < x_1 \le 1$, the corresponding sign pair in Table 3 is $(+,+)$: coexistence (in other words, protected dimorphism). Thus, the adaptive dynamics in Eq. (20) can drive the two branches to the boundaries, $x_1 = 1$ and $x_2 = 0$, without extinction of either branch.

What we clarify here is that in the case, $b_2 - c_2 < 0$, then, the dimorphic population undergoes bi-directional evolution that leads the levels of cooperative investment in the two branches to come closer and closer to each other. Different from the former case, the PIP in Fig. 3(e) indicates that for the two strategies given across the interior singular strategy $x = \hat{x}$, the possible sign pairs in Table 3 consist of not only $(+,+)$, but also $(+,-)$ and $(-,+)$. Thus, in the specific strategies, through the replicator dynamics, either of the two branches goes to extinction on the way toward converging to $x = \hat{x}$. This, however, does not matter for the evolutionary consequence. The resultant monomorphic population, whether it is from the former higher or lower branch, will continue evolving to $x = \hat{x}$. Indeed, the interior singular strategy $x = \hat{x}$ is convergence-stable for monomorphic populations.

This indicates that the continuous snowdrift game with $b_2 - c_2 < 0$, in particular with accelerating costs ($c_2 > 0$), will necessarily lead a traditionally differentiated population to converge to a monomorphic state with an intermediate level of cooperation (which is a local maximum).

**(d) How continuous snowdrift games affect social welfare**

We quantitatively compared evolutionary outcomes from the discrete and continuous snowdrift games. So far, we have calculated analytical expressions of non-trivial equilibria. Using the results, we accessed the quantitative difference of discrete and continuous strategies, which previously have only been discussed for matrix games (Tanimoto, 2007; Zhong et al., 2012).

First, we rewrite the difference in the cooperation levels at equilibria in Eqs. (13) and (18), as follows:

$$\hat{x}-\hat{n}=\frac{(b_2-c_2)(2b_2+b_1-c_2-c_1)}{2b_2(2b_2-c_2)}. \tag{21}$$

Then, the average payoff for the monomorphic population with an interior singular strategy $x=\hat{x}$ is given by

$$\bar{P}(\hat{x})=\frac{(b_1-c_1)(-4b_1b_2+3b_1c_2-c_1c_2)}{4(2b_2-c_2)^2}. \tag{22}$$

It should be stressed that maximal average payoffs in dimorphic populations, as well as in monomorphic populations, cannot be expected to predict the evolutionary outcome. In the discrete snowdrift game, at its interior mixed equilibrium $n=\hat{n}$ in Eq. (5), the average payoff over the population is given by

$$\bar{P}(\hat{n})=\frac{(b_2+b_1)(-b_2-b_1+c_2+c_1)}{2b_2}. \tag{23}$$

Indeed, our numerical investigations indicated that in specific parameters, the adaptive dynamics favor the second best equilibria, which bring about a lower level of average cooperation and/or payoff over the population (Fig. 5).

## 4. Discussion

So far, we have shown that the continuous extension of the well-known snowdrift game is more likely to lead to unification rather than diversification of cooperators and cheaters. We analyzed how allowing gradual evolution of cooperative investments can lead to outcomes that can qualitatively and quantitatively differ from discrete strategies. In the classical, discrete snowdrift game within well-mixed populations, the stable coexistence of cooperators and cheaters is a unique evolutionary outcome. Provided that the degree of cooperative efforts to produce common goods can continuously vary, however, this is often not the case. Indeed, we find that with a wider range of parameters (in particular in the case of accelerating costs) initially heterogeneous populations with high- and low-investment levels will be destabilized and merge into a homogeneous state in which all invest at the same, but intermediate, rate. Therefore, our analysis explicitly shows that the gradual evolution of cooperation often prefers the social inequity aversion in snowdrift games.

To describe intermediate levels of cooperation, an alternative and fairly trivial way to consider this is through the mixed strategies of C ( $x=1$ ) and D ( $x=0$ ) (McGill and Brown, 2007). In a mixed-strategy model it is assumed that a player invests 1 with probability $x$ , or otherwise, 0. It is known, however, that this treatment is structurally unstable (Dieckmann and Metz, 2006). We remark that the adaptive dynamics are linear with probability $x$ , which is identical to traditional replicator dynamics of frequency $n$ in Eq. (3), except for difference in the variables. Thus, it is obvious that invasion fitness at a singular strategy is completely flat: all strategies when rare can fit equally, corresponding to the results known by the Bishop-Canning theorem (Bishop and Cannings, 1978). Rare mutants can then sneak in along with the residents with a singular strategy by neutral drift, which yet is not predictable by the theory of adaptive dynamics.

It has also been considered that responding to disruptive selection can lead to sympatric speciation (Rueffler et al., 2006). By means of adaptive dynamics a mechanism for the disruptive selection to occur has become understandable as evolutionary-branching points (McGill and Brown, 2007). Interestingly, recent studies on speciation, by analyzing the empirical data, have clarified that for a specific kind of white fish, reversed speciation has happened in large European lakes, and thereby biodiversity has been greatly reduced (McKinnon and Taylor, 2012; Vonlanthen et al., 2012). Analogously, a mechanism for the reverse speciation to occur might be understood through the process of evolutionary merging. For instance, these studies of white fish indicated that species differentiation can be reversed by environmental eutrophication. Through the continuous snowdrift game, our analysis reveals that enriching the game environment, in particular the marginal benefit of cooperation in the population of cheaters, can increase the degree of $S-P$ in Eq. (1), and thus can reverse evolutionary branching, leading to an evolutionary merging of cooperation and cheating.

In previous numerical investigations of spatial snowdrift games, it was suggested that spatial coexistence does not always promote the evolution of cooperation (Hauert and Doebeli, 2004). Our results imply that the coexistence of cooperators and cheaters discrete in the structured population could be unstable when considering adaptive dynamics. Similarly, applying our approach to discrete games with more than two players or strategies (e.g., optional participation in public good games) deserves further investigation (Doebeli et al., 2004; Sasaki et al., 2015). More generally, evolutionary branching could be considered in the context of work specialization or cultural polarization (Axelrod, 1997; Torney et al., 2010; Valori et al., 2012). Evolutionary merging, can for example, suggest that a division of labor can be disbanded gradually, not abruptly, in a slowly changing environment.

On the one hand, our model has been minimalistic in that it only considers quadratic payoff functions. Considerable efforts looking at the evolution of cooperation among non-relatives, on the other hand, have so far clarified supportive mechanisms, such as direct or indirect reciprocity, reciprocity on networks, and multi-level selection, and promotion of cooperation in a heterogeneous population with cheaters (Rand and Nowak, 2013). Our results showed differences in the resultant cooperation level and average payoff in a case without such supportive mechanisms. Therefore, another fascinating question would be whether assimilation or dissimilation at the cooperation level would be a better environment that enhances social welfare when considering supportive mechanisms. This idea deserves further work, for instance, to explore whether evolutionary branching can facilitate the promotion of costly selective incentives in the presence of second-order free riders (Chan et al., 2013; Jiang et al., 2013; Xu et al., 2011; Xu et al., 2015).

**Acknowledgment.** T.S. was supported by the Austrian Science Fund (FWF): P27018-G11. T.S. also thanks to former financial support by the Austrian Science Fund (FWF): TECT I-106 G11 to Ulf Dieckmann at IIASA through a grant for the research project *The Adaptive Evolution of Mutualistic Interactions* as part of the multinational collaborative research project *Mutualisms, Contracts, Space, and Dispersal* (BIOCONTRACT) selected by the European Science Foundation as part of the European Collaborative Research (EUROCORES) Programme *The Evolution of Cooperation and Trading* (TECT). I.O. acknowledges support by Grants-in-aid for Scientific Research from the Japan Society for the Promotion of Science 22520160 and 26330387.

**Tables**

| | Conditions | Replicator dynamics for C and D | Title |
|---|---|---|---|
| I | $S-P>0$, $R-T>0$ | Unilaterally evolving to all C | By-product mutualism |
| II | $S-P<0$, $R-T>0$ | Bi-stable for C and D | Stag hunt |
| III | $S-P<0$, $R-T<0$ | Unilaterally evolving to all D | Prisoner's Dilemma |
| IV | $S-P>0$, $R-T<0$ | Coexistence of C and D | Snowdrift |

**Table 1.** Scenarios of replicator dynamics for discrete strategy C ($x=1$) and D ($x=0$).

| | Conditions | For monomorphism | For dimorphism across singular strategy $\hat{x}$ |
|---|---|---|---|
| i | $D(0)>0$, $D(1)>0$ | Unilaterally increasing to 1 | (No singular strategy) |
| ii-A | $D(0)<0$, $D(1)>0$ | Repelling from $\hat{x}$ | Converging to $\hat{x}$ |
| ii-B | | | Repelling from $\hat{x}$ |
| iii | $D(0)<0$, $D(1)<0$ | Unilaterally increasing to 0 | (No singular strategy) |
| iv-A | $D(0)>0$, $D(1)<0$ | Converging to $\hat{x}$ | Converging to $\hat{x}$ |
| iv-B | | | Repelling from $\hat{x}$ |

**Table 2.** Scenarios of adaptive dynamics for continuously varying strategy $x$ within $[0,1]$.

| | Conditions | Replicator dynamics for $x_1$ and $x_2$ |
|---|---|---|
| I | $S(x_2,x_1)>0$, $S(x_1,x_2)<0$ | Unilaterally evolving to all $x_1$ |
| II | $S(x_2,x_1)<0$, $S(x_1,x_2)<0$ | Bi-stable for $x_1$ and $x_2$ |
| III | $S(x_2,x_1)<0$, $S(x_1,x_2)>0$ | Unilaterally evolving to all $x_2$ |
| IV | $S(x_2,x_1)>0$, $S(x_1,x_2)>0$ | Coexistence of $x_1$ and $x_2$ |

**Table 3.** Scenarios of replicator dynamics for discrete strategy $x=x_1$ and $x=x_2$ with $x_1>x_2$.

## Figures captions

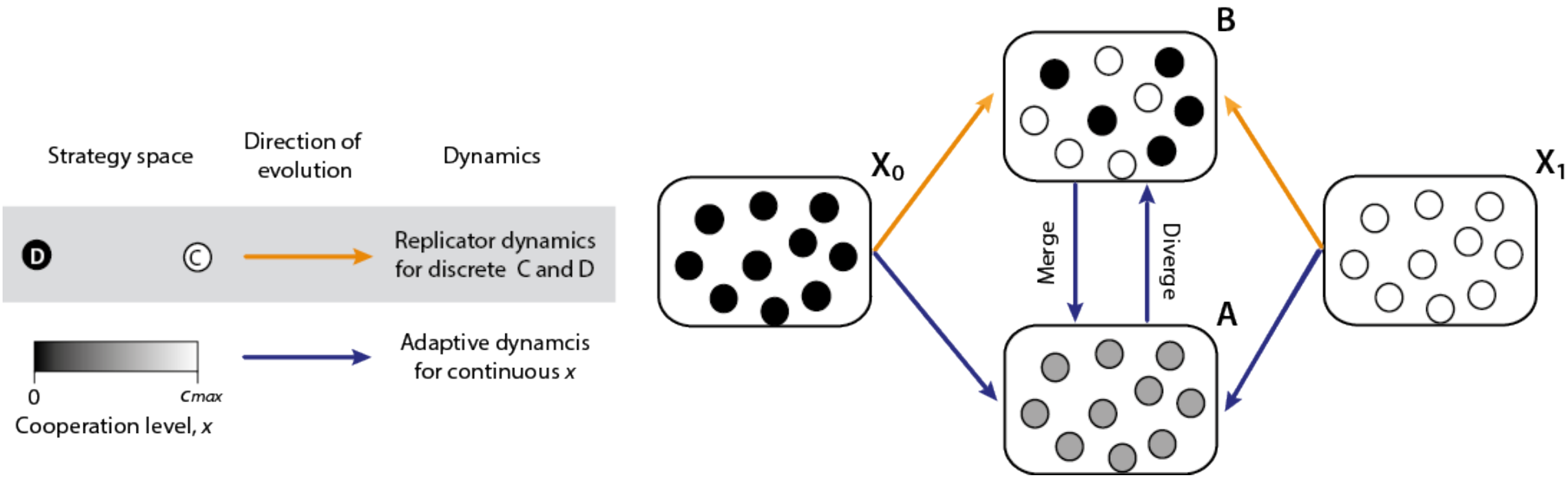

**Fig. 1.** Evolution of cooperation in snowdrift games. For discrete strategies, on the one hand, the evolution of the strategy frequencies can lead to the coexistence of cooperators and cheaters (upper arrows, $X_0$ to B and $X_1$ to B), yet do not help in understanding whether or not the resultant mixture is stable against continuously small mutations. For continuous strategies, on the other hand, the population converges to an intermediate level of cooperation (lower arrows, $X_0$ to A and $X_1$ to A) and can further undergo evolutionary branching (A to B). In this case, the population splits into diverging clusters across an evolutionary-branching point $x = \hat{x}$ and eventually evolves to an evolutionarily stable mixture of full- and non-contributors (B). Otherwise, it is possible that a point where $x = \hat{x}$ has already become evolutionarily stable. In this case, the initially dimorphic population across a point $x = \hat{x}$ can be evolutionarily unstable, and thus the population will approach each other and finally merge into one cluster at the point ("evolutionary merging").

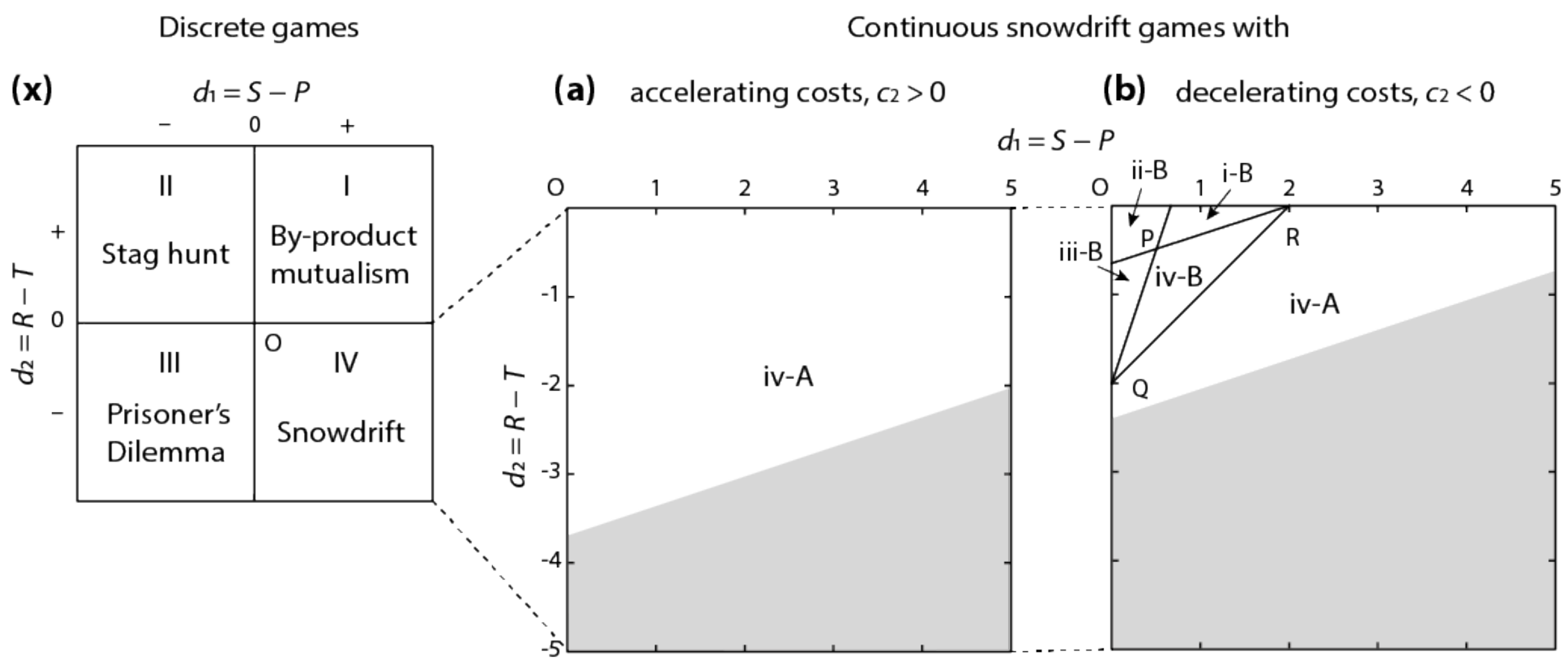

**Fig. 2.** Classification diagrams of evolutionary scenarios in snowdrift games. We employ $(d_1, d_2) = (R - T, S - P)$ as the coordinate system for parameterization. Parameter sets in the fourth quadrant, $\{d_1 > 0, d_2 < 0\}$, lead to the classical snowdrift game. However, parameters by which the diversified population of cooperators and cheaters can stabilize against continuously small mutations are restricted in the triangle OQR for decelerating costs $c_2 < 0$ (b), and do not exist for accelerating costs $c_2 > 0$ (a). Moreover, the sub-region for evolutionary branching to occur is sub-triangle PQR (iv-B). Compared to stabilization of the strategic diversity, its destabilization can happen within a wider region of parameters. Indeed, in region (iv-A) of (a) and (b), the mixed equilibrium in the classical snowdrift game is no longer stable under the continuous game. The two strategies will eventually converge to an evolutionarily stable state with an intermediate level of cooperation. In (b), these regions (iv-A) and (iv-B) are divided by line QR given by $b_2 - c_2 = (d_2 - d_1)/2 - c_2 < 0$. Lines PQ and PR are given by $D(0) = 0$ and $D(1) = 0$, respectively. In the shaded regions one of the natural assumptions, Eq. (18), does not hold: the benefit function $B(x)$ is not increasing. Parameters: $c_1 = 4.6$, $c_2 = 1$ (a) or $-1$ (b).

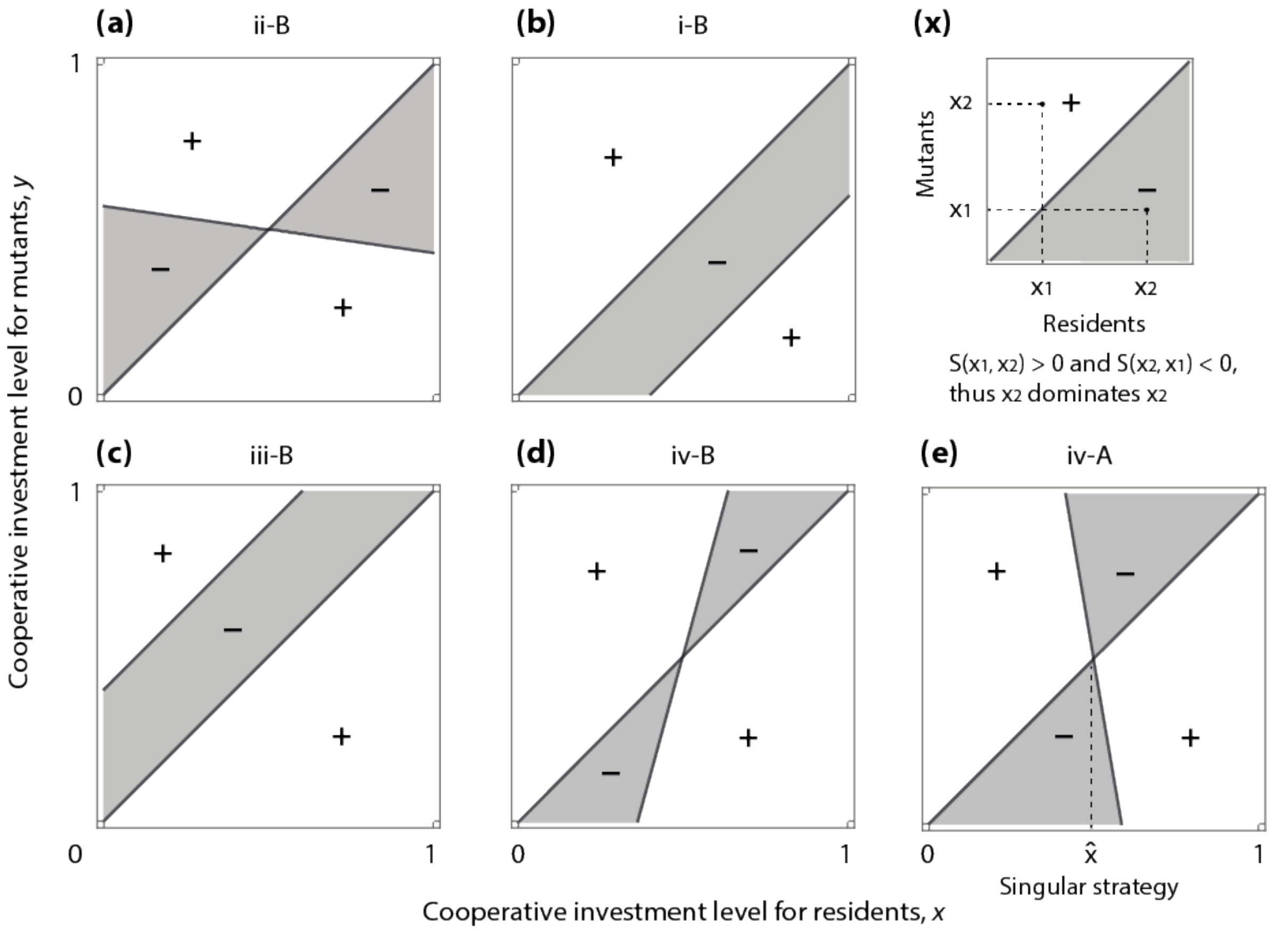

**Fig. 3.** Pairwise invisibility plots (PIPs) for the continuous snowdrift game. Each panel shows a sign plot of invasion fitness $S(x,y)$ in Eq. (11). Due to the linearity of the payoff difference with respect to the strategy frequency, the sign pair $(S(x_2,x_1),S(x_1,x_2))$ can indicate the frequency dynamics between the strategies with $x_1$ and $x_2$. Panel (x) exemplifies the case of $(S(x_2,x_1),S(x_1,x_2))=(+,-)$ which leads to a unilateral evolution: $x_2$ dominates $x_1$. The five sign plots are representative corresponding to the five cases of adaptive dynamics in the continuous snowdrift game: (a), (b), (c), (d), and (e) are for (ii-B), (i-B), (iii-B), (iv-B), and (iv-A), respectively. Parameters: $c_1=4.6$, $c_2=-1$; $(d_1,d_2)=(0.3,-0.3)$ for (a), $(0.7,-0.3)$ for (b), $(0.3,-0.7)$ for (c), $(0.7,-0.7)$ for (d), and $(1.7,-1.7)$ for (e).

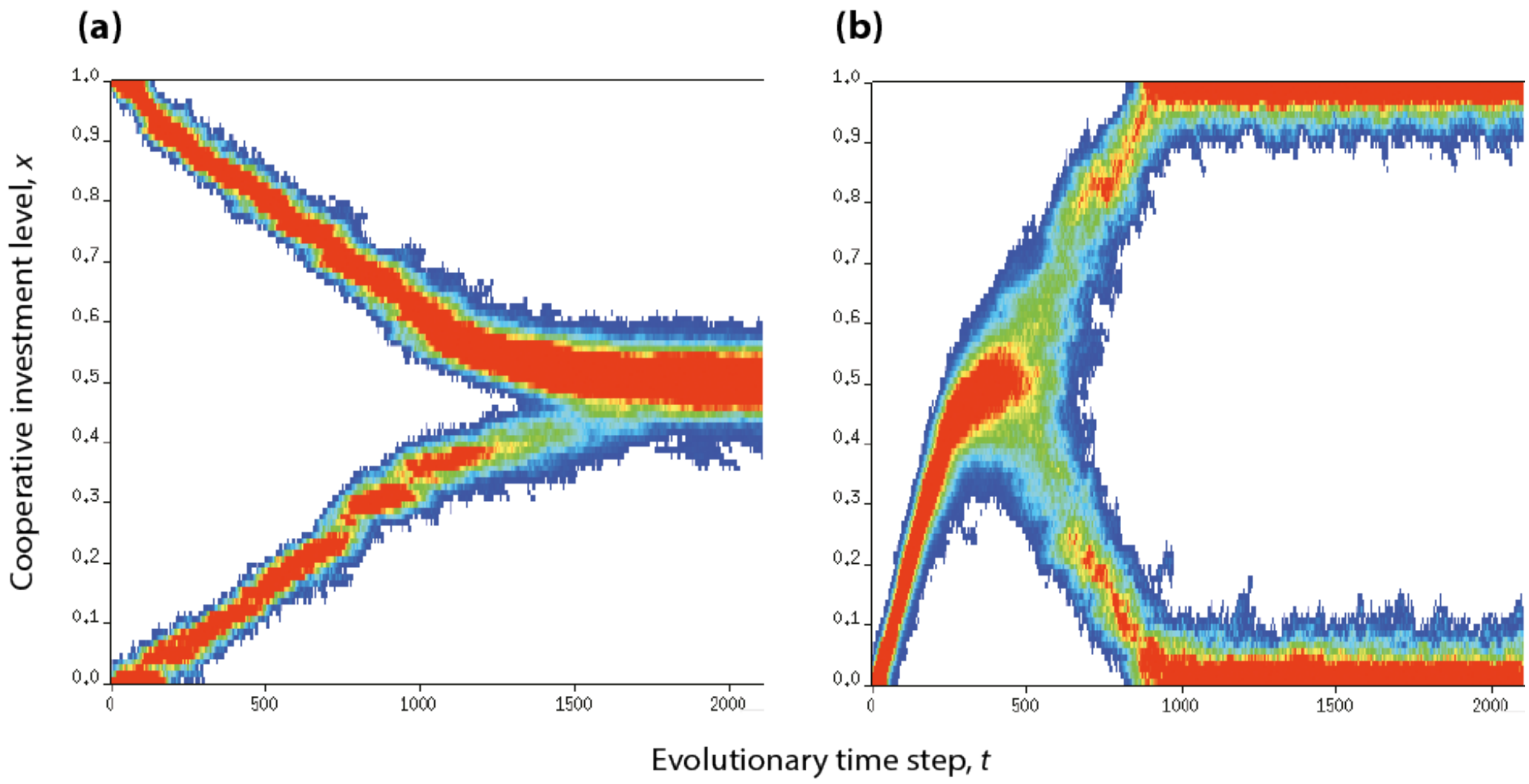

**Fig. 4.** Individual-based simulations of (a) merging and (b) branching in the continuous snowdrift game. Panels show evolutionary changes in the frequency distribution of cooperative investment levels over the population (from high to low: *red*, *orange*, *yellow*, *green*, *blue*, *white* (for 0)). At the outset of each tree, for (a) the population is at a traditionally acknowledged, mixed equilibrium with full-investment ($x = 1$) or non-investment ($x = 0$) and for (b) all have no investment ($x = 0$). In (a), the dimorphic population will eventually merge into a single branch. In (b), in contrast to this, the monomorphic population will first converge to an intermediate level and then diverge into double branches moving to the extreme states, respectively. Parameters: population size $N$ = 10,000, mutation rate $\mu = 0.01$, mutation variance $\sigma = 0.005$; for (a), $b_1 = 7$, $b_2 = -1.7$, $c_1 = 4.6$, $c_2 = -1$ ($d_1 = 1.7$, $d_2 = -1.7$); for (b), $b_1 = 6$, $b_2 = -1.4$, $c_1 = 4.8$, $c_2 = -1.6$ ($d_1 = 1.4$, $d_2 = -1.4$). In both cases the interior singular strategy is with $x = 0.5$. The scaling factor for proportional selection is set so as to be greater than the maximal difference over all possibilities of two samples.

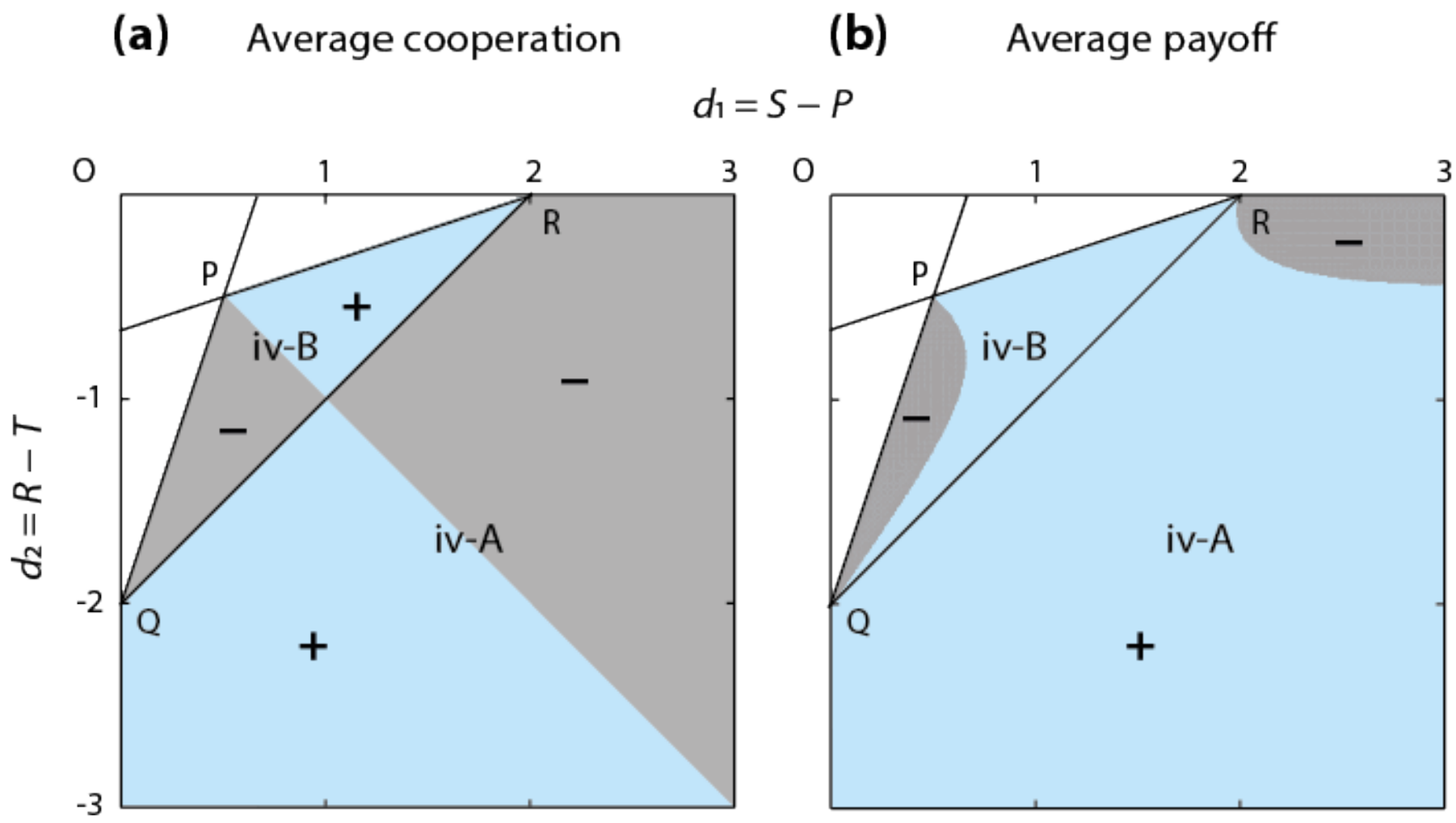

**Fig. 5.** Sign plots of differences in the average cooperation level and payoff over the populations with a classical mixed equilibrium with $\hat{n}$ in Eq. (18) and the interior singular strategy with $\hat{x}$ in Eq. (13). Parameters are as in Fig. 2. For each index, the sign is "$+$", if the value in the singular-strategy case is greater than that in the mixed-equilibrium case; otherwise, "$-$".